%% file: main.tex
\documentclass[conference]{IEEEtran}

\usepackage[numbers]{natbib}

\usepackage{fancyhdr}
\usepackage[utf8]{inputenc}
\usepackage{tikz}
\usepackage{amsmath}
\usepackage{listings}
\usepackage{fancyvrb}
\usepackage{dirtytalk}
\usepackage[frozencache=true,cachedir=minted-cache]{minted}
\usepackage{pgfplots}
\usepackage{pmboxdraw}
\usepackage{svg}
\usepackage[binary-units]{siunitx}
\usepackage{caption}
\usepackage{subcaption}

\usepackage{xurl}

\usepackage{inconsolata}

\usepackage{pifont}
\usepackage[hidelinks]{hyperref}

\begin{document}

\newtheorem{definition}{Definition}[section]

\providecommand{\keywords}[1]
{
  \noindent\textbf{\textit{Keywords}} #1
}
\date{}

\title{%
  Mapping Out the HPC Dependency Chaos
}
 

\author{%
  \IEEEauthorblockN{%
    Farid Zakaria\IEEEauthorrefmark{1}, %
    Thomas R. W. Scogland\IEEEauthorrefmark{2}, %
    Todd Gamblin\IEEEauthorrefmark{2}, %
    Carlos Maltzahn\IEEEauthorrefmark{1}} %
  \IEEEauthorblockA{%
    \IEEEauthorrefmark{1}University of California Santa Cruz, Santa Cruz, CA, USA\\
    \{\tt%
    \href{mailto:fmzakari@ucsc.edu}{fmzakari}, %
    \href{mailto:carlosm@ucsc.edu}{carlosm}%
    \}@ucsc.edu}
  \IEEEauthorblockA{%
    \IEEEauthorrefmark{2}Lawrence Livermore National Laboratory, Livermore, CA, USA\\
    \{\tt%
    \href{mailto:tgamblin@llnl.gov}{tgamblin}, %
    \href{mailto:scogland1@llnl.gov}{scogland1}%
    \}@llnl.gov}
}

\maketitle

\thispagestyle{fancy}
\lhead{}
\rhead{}
\chead{}
\lfoot{\footnotesize{
SC22, November 13-18, 2022, Dallas, Texas, USA
\newline 978-1-6654-5444-5/22/\$31.00 \copyright 2022 IEEE}}
\rfoot{}
\cfoot{}
\renewcommand{\headrulewidth}{0pt}
\renewcommand{\footrulewidth}{0pt}


\begin{abstract}

High Performance Computing~(HPC) software stacks have become complex, with the dependencies of some applications numbering in the hundreds. Packaging, distributing, and administering software stacks of that scale is a complex undertaking anywhere. HPC systems deal with esoteric compilers, hardware, and a panoply of uncommon combinations.

In this paper, we explore the mechanisms available for packaging software to find its own dependencies in the context of a taxonomy of software distribution, and discuss their benefits and pitfalls. We discuss workarounds for some common problems caused by using these composed stacks and introduce Shrinkwrap: A solution to producing binaries that directly load their dependencies from precise locations and in a precise order. Beyond simplifying the use of the binaries, this approach also speeds up loading as much as $7\times$ for a large dynamically-linked MPI application in our evaluation.
\end{abstract}


\input{text/introduction}
\input{text/taxonomy}
\input{text/discussion}
\input{text/shrinkwrap}
\input{text/related}

\input{text/conclusion}

\input{text/acknowledgements}


\bibliographystyle{IEEEtranN}
\bibliography{bibliography}

\end{document}

%% file: text/introduction.tex
\section{Introduction}

The Livermore Computing~(LC) facility at Lawrence Livermore National Laboratory~(LLNL) supports thousands of users on dozens of different clusters and hosts Sierra, ranked third on the Top500~\cite{top500} and is preparing for the upcoming deployment of El~Capitan~\cite{el-cap}. 
The simulation software written for these machines is complex and requires a large chain of dependencies that is continually growing.  In 2015, it was significant to say that some applications required 70 dependencies, counted as packages which may contain one library or dozens depending on the packager. Today the Axom~\cite{axom} library, a common support library for Livermore codes, can require more than \emph{200} total dependencies.
Each application and dependency can require specific compilers with specific runtime libraries and Message Passing Interface~(MPI) implementations to achieve the best performance, or to work at all.
Users may run several different codes or several different builds of the same code in the same environment as part of larger scientific workflows.
Immense effort has gone into building complex software pipelines reliably, such as Spack~\cite{gamblin2015spack} and EasyBuild~\cite{hoste+:pyhpc12} in HPC and in mainstream distributions as well, but far less work has gone into ensuring that the software will run consistently once it is built.

Part of the complexity in these systems comes from the fact that there is no single group or model in control of the package ecosystem on an HPC system.  They are instead composed of layers managed by different domains of responsibility.  As an example, a machine at LLNL will usually be based on RedHat Enterprise Linux~(RHEL) using the Tri-Lab Operating System Stack~(TOSS), an extension of RHEL, and on capability systems extended further by vendor extensions.  Most users will still not use compilers or much software from TOSS or the vendors directly, they use software from a separate site-specific development environment exposed by modules called TCE.  If it were only these four layers, the problem might be reasonably tractable, but even TCE only provides the bare essentials. Each group then manages their own software stack, or stacks, built on top of some combination of the lower layers and possibly other groups's manually managed stacks. These are managed without any common infrastructure or planning, either manually or with one of the HPC package managers or some combination thereof.

The challenges faced at Livermore with regards to managing software complexity are a microcosm of what exists in the wider software ecosystem. Anywhere that multiple maintainers and versions of software stacks coexist these problems do as well.
There has been an explosion in available software, packaging methodologies, and Linux distributions that compose them in an attempt to tame the chaos.
The pursuit of creating software with increasingly complex dependencies reproducibly and portably is becoming a defining problem at large across all segments of the packaging and distribution community.


The primary differentiator between how packages find their dependencies in different packaging methodologies is whether they rely on established conventions or enforced search paths on each binary.  The most common and well known method is to depend on a filesystem hierarchy similar to the FHS~(Filesystem Hierarchy Standard) to determine the components linked into a binary at runtime. Some more recent systems use other mechanisms to locate binaries, libraries, or other necessary portions of an application.  The first of these is the common mechanism used by nearly every Unix-like operating system and others throughout history.  Everything from Multics, Darwin, Haiku, and BeOS, to modern hermetic root systems like CoreOS, search for binaries and libraries at well-known paths in the filesystem. The other main group we refer to as explicitly linked, or store-based. The major differentiator is that libraries loaded, and directories searched, by a given application are determined explicitly as part of build and distribution in the system, rather than being implicitly defined by the conventions of the distribution.  This is accomplished in a number of ways and shows up in quite a few different places across the ecosystem. The trend toward containers like Podman or Singularity for distribution is one way to accomplish explicitly linking an application with all of its resources in a redistributable way.  It can also mean directly referencing target libraries in a binary using \Verb`RPATH` or \Verb`RUNPATH` as package managers like Spack and most manual HPC installation trees do. It can even mean explicitly controlling or patching the loader as is done by systems such as Nix~\cite{dolstra+:icfp08,dolstra+:lisa04} and Guix~\cite{courtes-guix-2015}.

All of these non-traditional distribution mechanisms, package managers, and techniques are working on mitigating an underlying issue:  The management of binary loading and interfaces is under-specified and managed all too often by pervasive arbitrary conventions.  Our goal in this paper is to discuss the various approaches used today and in the past, explore some of the issues and workarounds commonly found in HPC systems today, and to present Shrinkwrap, our tool implementing a workaround that gives us the ability to run a program with little need to adhere to these arbitrary conventions. We present the following contributions:

\begin{itemize}
    \item A survey of the state of practice in software distribution;
    \item A methodology, using existing loader mechanisms, to ensures binaries find their dependencies regardless of the user environment; 
    \item Shrinkwrap: A novel solution to caching dependency resolution for binaries, implementing the methodology;
    \item An evaluation of Shrinkwrap and use cases detailing its use at LLNL and problems it has resolved.
\end{itemize}


%% file: text/taxonomy.tex
\section{Common Practice of Software Distribution}

The taxonomy of packaging is synonymous with software deployment models.
The nuance and variation across deployment models are often
relegated to a small cohort across Linux distributions. Software
complexity and the proliferation of software has made software deployment,
specifically reproducibility, more relevant to the traditional user.

This section explores several common and rising software deployment models and discusses
the tradeoffs of each. It ends with a discussion of how these models are
composed to form the complete software ecosystem on an HPC system.

%
%
%
%
%

\subsection{The \emph{Traditional Model}: Filesystem Hierarchy Standard}

Established in 1994, and in continuous refinement since, is the Filesystem
Hierarchy Standard~(FHS)~\cite{fhs}, further extended with the XDG Base
Directory specification~(XDG)~\cite{xdg} in 2003 and the systemd
file-hierarchy~\cite{systemd-file-hierarchy} specification. This is by \emph{far}
the most common layout used in modern Linux distributions, and is so pervasive that it is frequently replicated as a component of the other models. This is the familiar base directories such as
\Verb`/bin`, \Verb`/etc`, \Verb`/lib` and others that we have all come to know.  
Using a consistent standard like this has huge advantages. Anyone can look at a system that is arranged
this way and determine where to look for things. Package managers that target this
model include some of the most venerable and well-tested in existence such as
Debian's apt and the rpm ecosystems.  These package managers work by
inserting software components into well-known locations on the system such as \Verb`/lib` or \Verb`/bin`. 


The goal of a single unified directory structure like this is curating a system as a unified whole. It forms a single coherent set of packages that work together seamlessly, and only rarely allows more
than one version of any given package.  Since all software components must reside within
a few key well-known directories, there is no easy way to support alternate
versions of a component beyond appending suffixes. This aids in updating software
components for security vulnerabilities, since there is only a single file that
needs to be updated, and the need for rebuilds and expensive updates is
minimized. Installation of a package is equivalent to writing files to
this single root one at a time, potentially overwriting existing files of
the same name. This multi-step approach to software delivery can leave the
system in an inconsistent state if the process is interrupted, especially
during distribution upgrades that replace critical base components like the C
standard library.  It is often difficult to undo a software deployment unless specific care
is taken to create backups beforehand.

\begin{figure}
\begin{tikzpicture}
\begin{axis}[
    ybar interval,
    xticklabel style={rotate=340},
    ymax=161332,ymin=0,
    xticklabels={Unversioned, Version Range, Exact},xtick={1,...,4},]
\addplot coordinates { (1, 141332) (2, 51316) (3, 17630) (4, 5)};
\end{axis}
\end{tikzpicture}
\caption{Debian package dependencies by type }
\label{fig:debian-deps}
\end{figure}
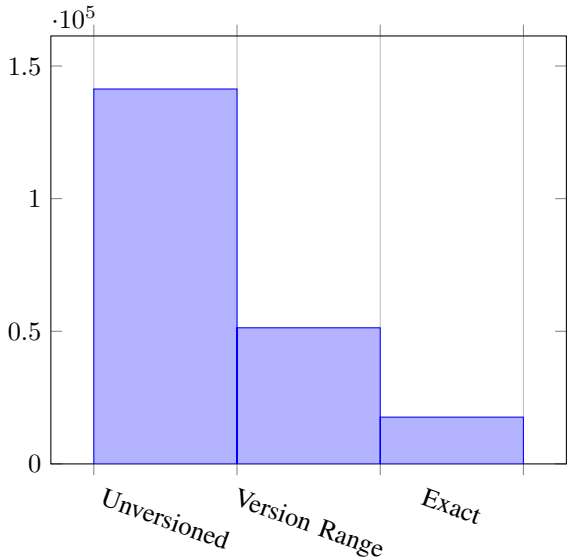

The lack of provenance of the files on disk is made worse by the
fact that most packages declare their dependencies without any explicit version
information. Figure~\ref{fig:debian-deps} shows an analysis of the Debian
package repository as of November 2021. Out of a total of roughly 209,000
packages, nearly $3/4$ of them use completely unversioned dependency
specifications.  These packages work because, and only because, the maintainers
of Debian diligently and manually ensure that the full graph of packages
in a given distribution build, link, and \emph{work} together.
That is an extremely impressive feat, but means that an immense amount of
knowledge about the needs of all of these packages is implicitly encoded and
unenforceable in software. This cost is also paid repeatedly by different
distributions such as Fedora~\cite{fedora}.  While some may benefit as downstream consumers of packages from
a parent distribution it is a fraction of the whole, leaving many distributions and package ecosystems duplicating the effort required to test, patch and compose packages.

\subsection{Self-Referential (Bundled) Model}

In an effort to make software applications more self-contained and easier to
manage, software components can be bundled along with their dependencies.  These
are typically vendored within the same directory or in a specially-marked
directory tree for the purpose. 
Bundled applications either rely
on a modified search path that prioritizes local libraries, normal on Windows or
Darwin and done by AppImage and AppDir formatted packages on Linux, or is replaced with a script that provides the appropriate options.
To use a wrapper script, the values are passed in a
variable like \Verb`LD_LIBRARY_PATH` to include the current working directory or a
subdirectory. 
The ELF header is also able to imitate this setup through the use
of the \Verb`$ORIGIN` expansion variable in \Verb`RPATH` and \Verb`RUNPATH`
entries to refer to the location of the binary. The tradeoff to this approach
is that there is a significant loss in the potential for deduplication across
binaries employing the technique.  Further, applying a security update for these
embedded
libraries will entail upgrading all of the bundles individually rather than
updating a central location.

The benefit is a far simpler experience for users. This is one of the main
reasons that Darwin applications can commonly be distributed in a \mbox{click-and-drag-to-install}
manner for example. It further provides a relatively simple linking model that allows an
application author to guarantee the libraries loaded are the ones they intended
rather than system libraries chosen by a downstream maintainer or a user.  The
software package can reside anywhere on the filesystem and is not subject to the
limited key space dilemma of FHS. Allowing multiple versions or minor
changes of the same software package to exist on the system at the same time
allows for the atomic installation and removal of the package as well. As a result, it is alluring for distributions of stand-alone scientific packages, especially graphical tools or others that are difficult to build and meant for novice end-users to install.

The challenge to this model is that it relies heavily on the software
developer. Catching build dependencies is challenging unless the builds are
always done in a sandbox environment. Because such a package uses its own vendored environment, it is very hard to extend, which poses significant challenges for packages that allow loading of user code or extensions.  For example, including an embedded Python interpreter in an application is straightforward, but should it support installing packages into that vendored python?  If so, where should they go? How should the user interact with them? It
can also pose a security issue because the user can choose where to place the bundle.  If the library path includes a writable directory, an attacker can leverage it to
load unintended code.

In terms of compatibility, this is not the most familiar model to Unix devotees,
but it is one of the best known to desktop users.  Beyond that, this model is
what allows almost all packages for Windows and MacOS to be installed without
extra dependencies outside of themselves.  The lack of external dependencies,
aside from the OS itself, is in fact so prevalent on Windows that the Microsoft
WinGet package manager \emph{does not support dependency specification or
resolution at all} as of December 2021~\cite{winget-cli-163}.  From a user
perspective, this model simplifies management and use but increases file size and
increases the chances of libraries going unpatched because they
are bundled with otherwise stable software that doesn't post updates to
dependencies.


\subsection{Hermetic Root Model}

\emph{Hermetic Root} systems are those whose goal it is to leverage many of the
same implicit assumptions as the FHS model while improving upon the
\emph{atomicity} and \emph{security} of the system. Given the pervasiveness of
the FHS model for software packaging and administration, these systems benefit
by providing a familiar and \mbox{easy-to-target} model.  The key insight they provide is
the creation of \emph{layers} in constructing the filesystem, similar to those
of \emph{overlayfs}, with the added ability to deploy layers via a commit model
that resembles \emph{git}. The ability to commit a new layer or rollback to
prior ones allows for the atomic delivery or rollback of installation or upgrade operations. The model
does not seek to impose any restriction on how the data is laid out, and adopts
any benefits or shortcomings of layouts used in addition to it. Although the
creation of filesystem layers and the curation of working packages still
represents a hurdle, once achieved the result can be distributed and reused
providing a reproducible environment with which to run desired workloads.  This
model also easily supports making the entire operating system, and vast majority
of package components, read-only with little effort making it resistent to 
many common classes of vulnerabilities.

The main challenge for this model in HPC is the slow move to support user namespaces in HPC centers.  As restrictions relax and more centers deploy bubblewrap with FlatPack and Singularity or Podman-based solutions, this may become more practical, but for now the lack of capabilities to create images inside of normal compute resources, and inside of security domains, makes this difficult to use on many HPC systems.


\subsection{Store Model}

The \emph{Store Model} refers to systems that install software components each
in an individual directory under a specific filesystem root, usually with each
individual package directory following the FHS. For example, Nix packages are
installed under \Verb`/nix/store` and spack packages under \Verb`<spack-repo>/opt/spack`.
This is a complete departure from the FHS model at the system level, moving
nearly all directories from a single root-level location to one per package.
References to dynamic libraries or shared code should only be done
from other store locations, and only explicit references in the package
descriptions should be respected. The explicit dependency linking between store paths
creates a directed acyclic graph of software components and their
dependencies. These systems often employ a consistent hash-naming scheme to
avoid conflicts, and allow arbitrary versions of the code to reside
congruently, providing the ability to perform upgrades or rollbacks
atomically by installing the whole new graph without invalidating the old one.
In order to exert control over the linking process, shared objects are resolved
by setting \Verb`RPATH/RUNPATH` during compilation, or through post-build actions
that modify binaries using \Verb`patchelf` or similar tools.


\begin{figure*}
    \centering
        \includegraphics[width=0.8\textwidth]{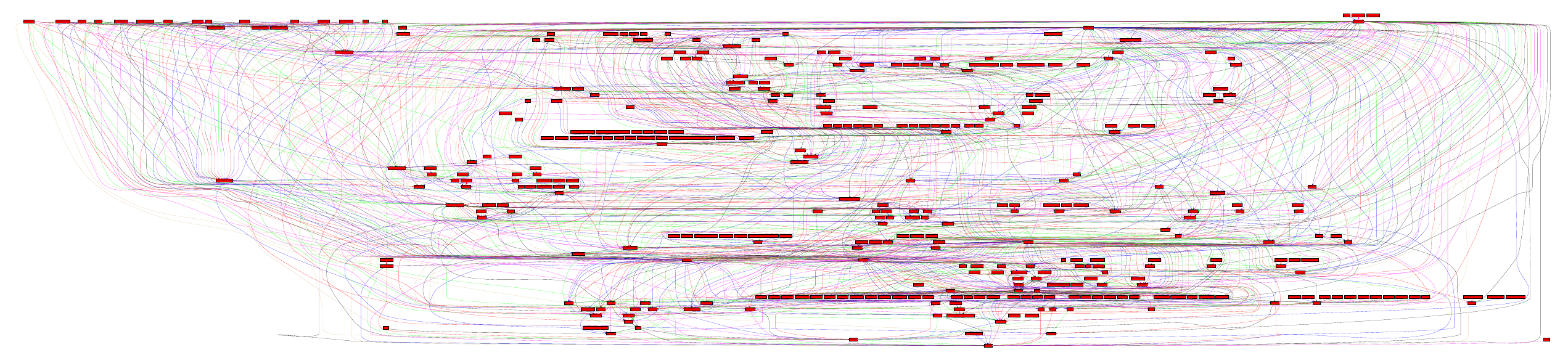}
    \caption{A graph, or snarl, of the build and runtime package dependencies needed by Ruby in Nix.}
    \label{fig:ruby}
\end{figure*}


Several new package managers and distributions employ this model, including Nix, Guix, and Spack.  These originated from concepts first introduced
through Nix~\cite{10.5555/1361397.1361410}. The model requires 
package authors to canonicalize the build steps into the system's model so that
the graph of all dependencies is explicit and complete. The consistent hashing scheme is
often referred to as \emph{pessimistic}, because it takes into account
the package's full source, build steps, and the same for its complete
transitive closure. Any minor change from source to compiler flags
for any package in the build graph will cause a domino effect of rebuilds.

This model fares well from an atomicity and reproducibility standpoint, but
security requires updates to propagate to all dependent packages, rewriting
potentially large segments of a system's packages when a popular library like
\Verb`libcurl` is patched. Even so, it is inherently no less secure than the
others since those updates are provided in a similarly timely manner, they just happen to be larger. Where this model runs into problems is compatibility, familiarity and implementation
given existing tools.  A NixOS system cannot natively run a dynamic executable
built on any other distribution even if the system has every single dependency
used by that executable.  There are projects that help deal with
this such as Nix-LD~\cite{nix-ld}, but the fact that everything in a Nix system is placed under
the store means that even fundamental building blocks like the loader~(\Verb`ld.so`) are not where an FHS system would expect them to be\footnote{In fact, Nix
patches away the ability for the linker to refer to default system locations or ld.so.conf}.  It is done for
good reason; this way a Nix system can use two different loaders with two C
libraries side-by-side without issue, but the compatibility is poor by default.

While we normally don't conceive of it this way, the development tools, distributions and module directories of HPC systems tend to be a manually curated version of a Store Model system as well.  This is hardly surprising, since that is a large part of the reason Spack builds the way it does, but when treated that way certain properties become clear.  For consistency and usability, it is desirable to make each package in such a module directory \emph{work} like a store model by ensuring that each package encodes all of its dependencies, rather than requiring environment variables loaded from a module.

Another issue that these models face is a model for loading of shared objects that was 
developed in and around the transition from SunOS to Solaris in the mid 1990s.
Limited to using \Verb`RPATH`, \Verb`RUNPATH`, or environment variables they tend to use a
combination of \Verb`RPATH` or \Verb`RUNPATH` and wrapper scripts to patch up the graph as
best as they can to produce a working system. However, these methods were never
meant to be used for this purpose, and create their own problems.  We'll discuss correctness issues further in Section~\ref{sec:discussion}, but loading performance can be significantly impacted by using a store model even when the setup is correct.  Since \Verb`RPATH` and \Verb`RUNPATH` only allow specifying paths to be searched, and they are all searched in order for each needed entry, applications with many dependencies end up searching many directories to find each library.  Figure~\ref{fig:ruby} depicts the dependency graph of the Ruby package in Nix with all 453 dependencies.  It is so dense, and so many components that it's nigh illegible, but it itself is a \emph{minor} dependency for many other packages. On a local filesystem, the overhead is usually small enough to be ignored, but when the dependencies are on a network filesystem it can be a significant performance issue.


\subsection{HPC and the Module Model}

We have eluded to the common setup of HPC systems in other parts of this section, but any given HPC system is usually comprised of layered instances of the FHS model and some form of the store model. Often the store model portion is less strictly structured and presented in the form of software modules handled by a module manager like \emph{lmod}. As an example, the software stack on Lassen (the open compute version of Sierra), the base system is an FHS formed from a combination of RedHat, TOSS and IBM base packages. On top of that, there is a large set of developer environment packages available through modules in \Verb|/usr/tce|. As of this writing, 338 separate directories are managed by application teams, many of which provide built versions of their software or tools for downstream consumers to use in whatever way they have decided in their own tree.

Within any one component of this system, packages tend to be self-consistent, usable and stable.  Difficulties arise from combining elements of multiple components, and the strategies used to compose them.  Common issues include: one layer using \Verb|RPATH| to ensure all dependencies can be found while another uses \Verb|RUNPATH| which causes the \Verb|RPATH| to be ignored; runtime libraries injected by compilers without \Verb|RPATH| entries added relying on the environment; and difficulty identifying which packages are ABI compatible with one another or which compilers use which runtime library versions. 
Applications are composed from some combination of these components, and frequently more are pulled from package managers like Spack, vcpkg, pip, conda, as well as other sources.  The chaos of these deployment strategies, tools, and requirements can confound even sophisticated application developers and dedicated HPC support teams, resulting in fragile software deployments.

%% file: text/discussion.tex
\section{Discussion}
\label{sec:discussion}

The survey of packaging methodologies demonstrates that different means of
bundling software within the Linux environment are possible and can achieve
varied levels of atomicity, reproducibility, and security. In each classification,
the system relies on a shared set of simple primitives controlling the loader to
distinguish itself. These differences are subtle to all except the most well-versed
in the space. For instance, the \Verb`RPATH` specified within the ELF header has
precedence over all dynamic loading search locations \emph{unless} \Verb`RUNPATH` is
set, in which case it is \emph{ignored}. Additionally complicating one's understanding of
how libraries are resolved is that \Verb`RPATH` entries in each ancestor are searched,
whereas \Verb`RUNPATH` entries are not.

\subsection{Issues with {\tt RPATH} and {\tt RUNPATH}}
\label{sec:issues}

Understanding the link order of a binary is difficult in practice. Are needed libraries traversed breadth first or depth first? Using what mechanism are the libraries
ultimately resolved? These questions may seem innocuous, but they are critical to what components load in a given environment.
Issues can cause the link order, and thus the loaded dependencies, to subtly change
long after installation and testing.

To answer some of these questions, mostly in the context of the glibc loader, shared objects are only loaded into memory a single time
during traversal, usually based on their \Verb`soname`. If a shared object has already
been visited and is needed by another dependency it will be provided without a
lookup, and will not raise a warning or error if that library would not have
been found otherwise. That is a useful performance optimization, but it also
allows missing path entries to hide in working binaries that may surface later
when the binary is run with a different set of flags or a new version of a
library in the tree.  Listing~\ref{lst:missing-runpath} shows an example of a
library trace from a program called \Verb`dbwrap_tool` where the application and
many of its libraries use \Verb`RUNPATH` to find what they need, but one library four
levels down the tree has no \Verb`RUNPATH`.  The \Verb`libsamba-modules-samba4` library
finds three of its dependencies through default search paths, but the fourth
wouldn't be found at all if it hadn't been loaded earlier in the tree by another library with a correct \Verb`RUNPATH`.

\begin{table}[htpb]
   \centering
   \caption{Properties of {\tt RPATH} and {\tt RUNPATH}}
   \label{tab:properties}
   \begin{tabular}{r|l|l}

Property    & {\tt RPATH} & {\tt RUNPATH} \\
\hline
Before {\tt LD\_LIBRARY\_PATH} & Yes   & No      \\
After {\tt LD\_LIBRARY\_PATH}  & No    & Yes     \\
Propagates  & Yes   & No      \\

   \end{tabular}
\end{table}

\begin{listing}[htpb]
\begin{minted}{console}
$ libtree /usr/bin/dbwrap_tool 
├─ libpopt-samba3-samba4.so [runpath]
│  ├── libcli-smb-common-samba4.so [runpath]
│  │  ├─ libiov-buf-samba4.so [runpath]
│  │  ├─ libsmb-transport-samba4.so [runpath]
│  │  ├─ libsamba-sockets-samba4.so [runpath]
│  │  ├─ libgensec-samba4.so [runpath]
│  │  │  ├─ libsamba-modules-samba4.so [runpath]
│  │  │  │  ├─ libsamba-util.so.0 [default path]
│  │  │  │  ├─ libtalloc.so.2 [default path]
│  │  │  │  ├─ libsamba-errors.so.1 [default path]
│  │  │  │  └─ libsamba-debug-samba4.so not found
├─ libdbwrap-samba4.so [runpath]
├─ libutil-tdb-samba4.so [runpath]
├─ libsamba-debug-samba4.so [runpath]
├─ libpopt.so.0 [default path]
├─ libtalloc.so.2 [default path]
├─ libsamba-errors.so.1 [default path]
├─ libsmbconf.so.0 [default path]
└─ libsamba-util.so.0 [default path]
\end{minted}
\caption{A demonstration that binaries can work due to shared objects being
found by searching earlier paths}
\label{lst:missing-runpath}
\end{listing}

\begin{figure}
    \includesvg[width=\columnwidth]{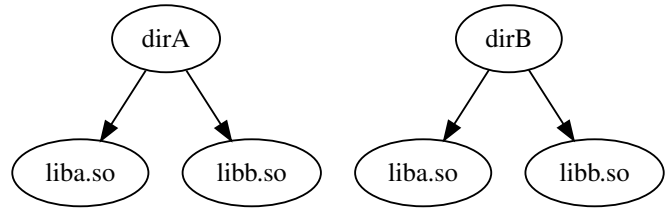}
    \caption{A paradoxical setup for \Verb`RUNPATH` where the desired libraries are
    {\tt dirA/liba.so} and {\tt dirB/libb.so}}
    \label{fig:paradox}
\end{figure}

The use of \Verb`RUNPATH` to instruct the linker, such as in the \emph{Store
Model}, is potentially problematic given that the granularity of the search path is at the
directory level. Without a direct mapping of needed shared objects to their
respective location, it is possible to find an incorrect object. The usual
solution to this is to ensure that the order of items in the search path
gives the correct result, but even simple conflicts can produce cases where this
is no longer possible.  Consider a system with libraries arranged as in
Figure~\ref{fig:paradox}, in which \Verb`liba.so` is needed from \Verb`dirA` and
\Verb`libb.so` is needed from \Verb`dirB`. In any ordering of any of the available search
path options, there is no way to get the correct intended behavior without creating a new directory with the correct versions.

These issues have inspired a great deal of debate on the use of any of these
mechanisms.  The Debian community has publicly argued over
policies in this space for many years; some of the resulting outcomes are now
documented on their wiki~\cite{debian-rpath}.  The main concern centers
around an application loading different versions of a library depending on
loader order, arguing that the dynamic linker should solve this rather than
\Verb`RPATH` or \Verb`RUNPATH` entries.  That seems to be arguing for use of \Verb`ld.so.conf` or
similar to resolve these issues, which makes sense from the perspective of a
distribution maintainer who can control that configuration easily and desires a single coherent FHS system.  Some of the caution also comes
from early use of \Verb`RPATH` by tools like \Verb`libtool`, which once included
directories that also existed in \Verb`ld.so.conf` in the \Verb`RPATH`, locking system paths
into the search out of order.  The general desire to avoid \Verb`RPATH` is not new. In
fact, the first version of the manpage for \Verb`ld.so` on Linux~\cite{ld.so-manpage} included a statement that
\Verb`RPATH` was deprecated and should be avoided.

On the other side, the Qt project published a recommendation~\cite{qt-rpath}
that authors of Qt applications should use \Verb`RPATH` and ensure \Verb`RUNPATH` isn't set.  Given
the distaste for \Verb`RPATH`, that may seem surprising, but Qt loads component
libraries from code inside other libraries.  As a result, an application that
uses QtGui with an \Verb`RPATH` will provide its \Verb`RPATH` entries to the load issued in
QtGui, allowing the system to find the correct version of the dependency.  If
the application has any \Verb`RUNPATH`, the search will only include those paths set on
QtGui itself.  Normally this is even treated as a feature rather than a bug, and the
QtGui library should have a \Verb`RUNPATH` to find the dependency.  That works as long
as the dependency tree is strictly a tree, but if the application provides a
plugin or a library that gets loaded, say by \Verb`dlopen`, in a different shared
object, it has no way to provide search paths to that dlopen with \Verb`RUNPATH`.  The
only recourse at that point is \Verb`LD_LIBRARY_PATH`, which can cause even more
damage in a \Verb`RUNPATH`-based system where more of the paths can be overridden in
sub-applications, or adding an explicit API to load the library by absolute path.

From an administrator perspective, working with either \Verb`RUNPATH` or \Verb`RPATH` in an
executable or library can cause pain points.  If a library is locked to point to
a library at \Verb`/opt/rocm-4.3.0` and that version is found to be buggy but binary
compatible with 4.3.1 for example, then replacing 4.3.0 with 4.3.1 is more
costly than without those paths set. They are forced to either recompile the
library, symlink the new one into an inappropriate location, or override at the
application level with \Verb`RPATH` or \Verb`LD_LIBRARY_PATH`.  If the library uses \Verb`RUNPATH`,
the \Verb`RPATH` solution on the application is insufficient as well, and can cause
situations where parts of one version and parts of another are loaded due to two
different search path orders with respect to \Verb`LD_LIBRARY_PATH`.

In the end, traditional search path management is fragile and poorly suited to
store-based or module style package distribution, or separating packages by
directory in general.

\subsection{Questioning Dynamic Linking}

\begin{figure}
    \centering
    \includesvg[width=\columnwidth]{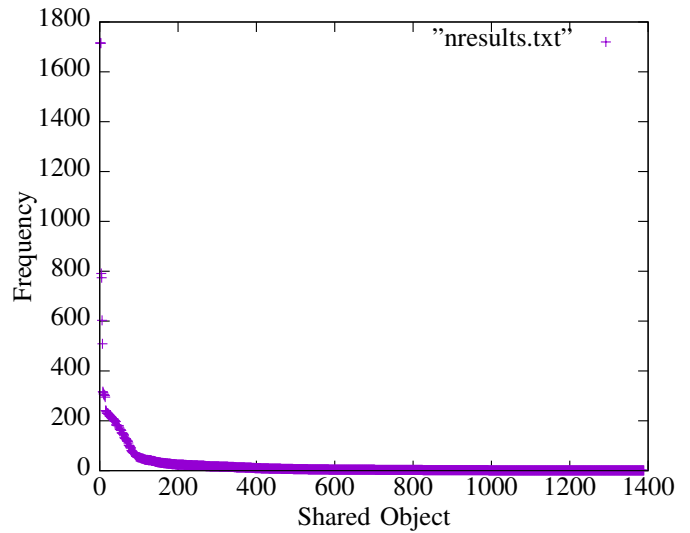}
    \caption{Shared object reuse on a typical Debian installation with 3287 binaries. Only ~4\% of shared object files are used by more than 5\% of the binaries}
    \label{fig:object-reuse}
\end{figure}

The necessity for package curation in distributions is a point of contention
as many language package managers trend towards vendoring dependencies to help improve reproducibility
of the package~\cite{fedora-golang}~\cite{fedora-deps}. What at first became a solution to managing \emph{dependency hell}
soon became a \emph{holy grail} to manage a single-unified graph of all software packages. The need for
dependency-shared object resolution arose from a time when storage and bandwidth were expensive and in limited
quantities or availability. The ability to upgrade a buggy or vulnerable package with as small as a change as necessary to the dependency graph
is a useful requirement of the past. There has been ongoing public discourse~\cite{devault-dynlib}~\cite{static-harmful}
that demonstrates the total cost to re-download \emph{all binaries} affected by CVEs in 2019 to be under \SI{10}{\gibi\byte} (significantly smaller if you discount glibc).
A survey of a local machine with 3,287 binaries demonstrates that the majority of libraries are used by relatively few binaries on the system as demonstrated in Figure~\ref{fig:object-reuse}. Why is dynamic linking continued given the lack of shared object reuse?

Much of this paper has been focused on the pitfalls and short-comings of dynamic linking, many of which
are non-existent for a statically compiled executable. That said, in an HPC context, we have seen leadership class systems with only static linking that deduplicated statically linked binaries in memory, and a transition back to supporting dynamically linked binaries.  The memory reuse benefits can be more noticeable when running the same application as one process per core and the same set of libraries loaded.  In the wider community, Linux distributions have surfaced to explore the emerging pro-static philosophy~\cite{stali}, but have yet to gain meaningful traction.

Dynamic loading also provides one significant benefit we have not otherwise discussed.  Many tools, especially prevalent in HPC, rely on dynamic linking to override or wrap symbols.  For example, tools that use the PMPI interface are usually pre-loaded with \Verb|LD_PRELOAD|; same for performance or memory tools like gperf.  Changing to fully static linking breaks \emph{all} of these tools, rendering them unusable.  That may still be worthwhile for final production release binaries in some cases, but it means that a great deal of work during development may need to use dynamic binaries.

\subsection{Questioning the Loader Interface}

In the mid-to-late 1990s, the tools provided by the linker allowed augmenting the search space for shared
objects in well-intentioned ways, but these techniques have not aged well.  \Verb`RPATH` was sufficiently misused and thus reviled on
SunOS, Solaris, and later on Linux that it has been deprecated for 20 years. Yet,
it is still supported and in active use in newly written systems to this day.
Its replacement, \Verb`RUNPATH`, was meant to solve developers' issues by ensuring
that \Verb`LD_LIBRARY_PATH` could override search paths in binaries, and that executables would not
pollute their dependencies with their embedded search paths.  While it succeeded in these goals, by not disallowing \Verb`RPATH` and \Verb`RUNPATH` from being used together and by eschewing propagation, it lost the
ability to propagate search paths to dependencies when necessary.

The constraints we want to express are a combination of options
to inject new paths into the library search path:
\emph{prepend, append, and whether to inherit}. All but one of the problems listed in
Section~\ref{sec:issues} can be solved by offering prepend/append and a
boolean propagation flag on each path added to the search space.

The desire to find all libraries within a few well-known directories from the
FHS model has bled into the ability to specify the search space as a single
list for all libraries.
Allowing
the ability to dictate the search space per shared object would give fine-grained
control over the search semantics.  This would also solve the final
issue: the ability to load libraries with conflicting filenames from paths
deterministically.

Having considered this direction, the other question that must be asked is why
we must rely on a bare soname and a linear set of paths to search to do this
job.  The Fuchsia kernel and Zircon system loader implement a service
to request dynamic libraries at load time, allowing load configurations to be
changed between libraries during loading~\cite{fuschia-loading}.  In the end,
though, there is a standard ELF loader on top of it.  Given the option to change
the way dependencies are encoded in binaries could allow a system like Nix or
Spack to store the hash of the library being requested, store the specification used to
build it, or store enough information to be able to not just load it but
determine with far greater detail which version is expected if it is not
available.  One can envision a system that would allow a user to take a binary
set up that way and ask a tool to provide all of the dependencies it needs in place
of distributing a static binary or a container.

\subsection{Workarounds}

While it's a pleasant thought experiment to imagine a world where we do not require backwards compatibility with established
loaders, the state of the practice is that we must work within the limitations of ELF and the System V ABI model.
Therefore, we propose a number of workarounds to allow packaging models to avoid
some of these issues given the current capabilities of loaders.
Additionally, we will discuss what seem to be the core features desired in a
future loader to allow all of the package author, package maintainer, and the
end user to request the behavior they want without requiring massively long
search paths and ambiguous lookups.

\subsubsection{Dependency Views}

This method is based on the concept of environment views from the Spack package
manager, but could be used by any system that wants to do per-package resolution
of dependencies. Rather than setting \Verb`RPATH` or \Verb`RUNPATH` entries on the executable
and every library to all dependencies, each gains a single \Verb`RPATH` or \Verb`RUNPATH` to a
package-local directory containing an FHS-styled filesystem populated with
symlinks to the package's dependencies. Rather than a long list of \Verb`RPATH`s, there
is now only one, and resolution should necessarily be much faster, especially on
network filesystems where stating a file can be slow.  An extra benefit to this is that it may work for resources other than dynamic libraries such as data or font packages, by providing the combined view of those dependencies a package may expect from a traditional FHS system.  The downside is that this
method requires both a tremendous number of symlinks, and thus filesystem inode
resources, to represent a full system.  It is also constrained to only allowing a
package to depend on a single version of any dependency, since they cannot
link on top of each other.

\subsubsection{Needy Executables}\label{needy-executables}
A less costly workaround in terms of filesystem resources is to use
the issue shown in Listing~\ref{lst:missing-runpath} to our advantage.  Since
libraries are cached by soname, and libraries are loaded in breadth-first-search
order starting from those needed by the executable, we can fix the load order in the executable.  We do this by by directly linking all libraries required by the full transitive closure of dependencies into the executable.
For example, consider an \emph{executable} that depends on
\Verb`liba` which depends on \Verb`libb`. Rather than having a normal needed list of just
\Verb`a`, the binary would instead have a needed list of \Verb`a,b` and \Verb`RPATH` or \Verb`RUNPATH` entries to find both.  The \Verb`RUNPATH` issue with propagating search paths to dependencies is
partially mitigated here as well, since libraries are pulled to the top for
resolution.


Despite working around some of the main issues of the traditional approach, this
method still has flaws.  If any pair of libraries in the set define the same strong symbol, the link will fail.  Additionally, load paths for
\Verb`dlopen` calls without a path are not directly resolved by this method.  Since these are loaded programmatically,
they are not part of any needed entry.  We could envision a system that traces
all such calls and adds the libraries to the needed list, but that could cause
breakage due to initialization order or load parameters such as the local or
global nature of symbols in the library.

%% file: text/shrinkwrap.tex
\section{Shrinkwrap}

The introductions of Spack and similar store-like systems have added a much
needed level of reproducibility and organization to managing the common
combinatorial stacks required by the HPC community. The ability for a
binary or shared object file to explicitly tell the dynamic linker through the
use of \Verb`RPATH` to search specific content-addressable named locations is
pivotal to this approach, as it is to the manual approaches described earlier.

Unfortunately, the ability to modify a dynamic linker's search algorithm is very
coarse. All of \Verb`RPATH`, \Verb`RUNPATH`, and \Verb`LD_LIBRARY_PATH` are simply lists of directories, and apply to all dependencies needed by
the binary. As the number of dependencies for a shared object grows, so does
the length of the list that must be searched, penalizing the startup time for
the process. This \emph{unnecessary} work has real consequences.
\citeauthor{10.1145/2464996.2465020} have written how with sufficiently large
dependency graphs, one can flood the filesystem with requests and have process
startup times on the order of hours~\cite{10.1145/2464996.2465020}.

Based on our experiences, we have developed \emph{Shrinkwrap}, an
open-source implementation of the \emph{Needy Executables} option presented in
Section~\ref{needy-executables}. Shrinkwrap provides the following features:

\begin{itemize}
  \item Encodes dynamic dependencies in the binary by their absolute path;
  \item Lifts all transitive dependencies to the top shared object to simplify
        auditing and prevent \Verb`RPATH`/\Verb`RUNPATH` interference in transitive dependencies;
  \item Offers virtual resolution strategies to handle cross-platform
        binaries or alternative dynamic linkers;
  \item Available as open-source MIT licensed software.
\end{itemize}

When faced with a recurring problem, often the solution is to cache the previous
answer to avoid unnecessary work. Shrinkwrap adopts this approach by
\emph{freezing} the required dependencies directly into the \Verb`DT_NEEDED`
section of the binary.  Rather than listing the soname each entry is an absolute path. Furthermore, the transitive dependency list is lifted to
the top-level binary to simplify auditing the required dependencies. All of the
needed dependencies, including transitive dependencies, are now listed by
absolute paths on the top-level binary. Shrinkwrap is written in Python,
leveraging the lief~\cite{lief} library for parsing and writing ELF binaries.
Lief was chosen for its clean interface, ability to work stand-alone, support for
symbol analysis, and an option to support MachO and PE binaries to offer
similar benefits on MacOS and Windows in the future.

\begin{figure}[h]
    \includegraphics[width=\linewidth]{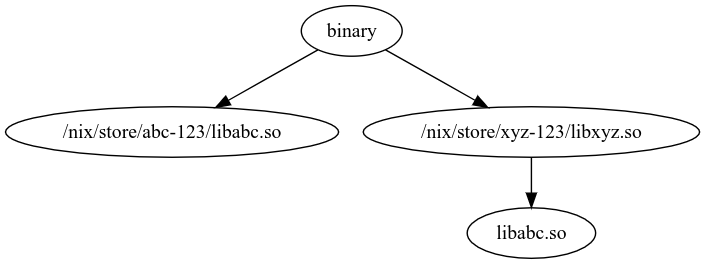}
    \caption{Deduplication based on soname}
    \label{fig:soname-dedupe}
\end{figure}

The solution is conceptually simple, but applying the modifications to the
binary in a general fashion is challenging.  Our desire is to support all
reasonably compliant linkers and loaders on Linux. In practice, Shrinkwrap
currently supports glibc binaries and others that use the same loader behavior
such as BSD libcs, but not musl~\cite{musl}.
Shrinkwrap relies on the dynamic linker deduplicating libraries
with a common file basename or whose \emph{soname}~(ELF header value) are the
same. Consider the example in Figure~\ref{fig:soname-dedupe}:
Shrinkwrap elevated \Verb`libac.so` to a direct absolute dependency of the
binary, but relies on the dynamic linker deduplicating the resolution for
\Verb`libxyz.so`, which does not refer to it absolutely. Referencing
dependencies by their absolute path makes it impossible to
swap out dependencies for alternative libraries using traditional methods like \Verb`LD_LIBRARY_PATH`.
The use of \Verb`LD_PRELOAD` remains viable, so in cases where specific functionality would still be preferred
to be overwritten, a backdoor into dynamic linking remains. This also means that traditional preloaded tools continue to work as normal.

When implementing Shrinkwrap against \emph{glibc}, the deduplication is
performed and the necessary libraries are resolved correctly. Unfortunately,
other dynamic linkers such as \emph{musl} do not exhibit the same behavior,
causing the solution to not be compatible across other environments.  The musl
loader does not cache libraries loaded by their full path by soname, but by inode number, causing
some load order issues with our scheme.  They also do not implement the standard
behavior of either \Verb`RPATH` or \Verb`RUNPATH`, but a meld of the two where
paths are inherited by dependencies but are searched \emph{after}
\Verb`LD_LIBRARY_PATH`.  This behavior would actually solve a number of problems
with \Verb`RUNPATH`, but since it is non-standard it makes supporting musl more
difficult for a tool like Shrinkwrap.  This incompatibility was raised to the
musl developers on their mailing list~\cite{musl-mailing}, however, the primary
challenge is that while the System V ABI specification requires dynamic
libraries to be deduplicated, or at least not to be loaded redundantly, it
\emph{does not} specify how they are to be matched to determine if they are
duplicates.

Aside from dealing with divergent loader behaviors, Shrinkwrap must also identify
which library on the filesystem each needed entry resolves to.  In a simple
case, using Shrinkwrap on the target system, we can use \Verb|ldd| or run the
binary interpreter extracted from the binary with an option to list, as in
\Verb|ld.so --list|, to get the actual behavior the loader would use given
current conditions.  When that works it gives Shrinkwrap exactly what we need
and ensures consistent behavior.  To handle cases where binaries are not
executable on the current system, or where the loader is either not usable in
this way or not executable itself, Shrinkwrap also offers a native strategy that
traverses the filesystem the way that the loader would to find libraries.  This
is a useful option, but the number of corner cases is large.  Mainly the issue
is that the System V standard says libraries that do not match the architecture
of the loading binary should be silently ignored, so we must detect these and
avoid them since they are very common on systems with multiple native
ABIs (x86 and x86\_64 for example).  Additionally, glibc supports loading more
specialized versions based on the target architecture from subdirectories of
each directory in the search path, and other expansions which must be faithfully
replicated.

Shrinkwrap works strictly by replacing the \Verb`DT_NEEDED` libraries of the binary,
however, dependencies may still be resolved via additional means such as \Verb`dlopen`. Binaries whose
runtime dependencies outnumber their static dependencies may not see optimal improvements.
An area of future work as outlined in Section~\ref{needy-executables} would be to allow Shrinkwrap to audit
all \Verb`dlopen` calls and lift them as \Verb`DT_NEEDED` so they can be easily referenced by absolute path.  For cases where the user or packager knows what libraries will be dlopened, and the semantics allow, adding the names of these libraries to the needed section before using Shrinkwrap allows Shrinkwrap to resolve them as well.  This works well for python modules for example, since they load cleanly and don't init until called.  When the libraries are unknown however, perhaps plugins that aren't even installed in the same package, we will consider other mechanisms as future work.

All of that said, our tests across a varied array of binaries from Nix, Spack
and hand-built HPC codes on a variety of architectures have given us confidence
that Shrinkwrap is capable of handling most binaries found in the wild.  Some
example applications are discussed as part of Section~\ref{sec:evaluation}.

\section{Evaluation of Shrinkwrap}
\label{sec:evaluation}

In order to evaluate Shrinkwrap as an approach to resolve practical issues with dynamic
loading, we present evaluations of both the performance characteristics of a
shrinkwrapped binary and case studies of applying Shrinkwrap to difficult library
resolution problems. The performance of executing Shrinkwrap itself is bounded mostly by
the time to traverse the filesystem, and if necessary, rewrite portions of the binary.
To wrap a binary with 900 needed entries and an \Verb`RPATH` 900 entries long with a 213MiB
main executable, took either four seconds on a Xeon E5-2695 system with the filesystem
cache warm, or over a minute on a cold NFS cache. Since the operation is intended to be
done only rarely, and usually on much smaller applications, its performance is
sufficient. More important is the loading performance of binaries that have been
shrinkwrapped, and the improvements in ergonomics of creating and using binaries on
complex systems.

\subsection{Loading Performance}

The number of library dependencies needed by a particular binary and the number
of entries in the \Verb`RUNPATH` can vary greatly. Consider a highly dynamic but common
binary, the \emph{emacs} editor,
as built by Nix, lists {\bf 36} directories in its \Verb`RUNPATH` and requires
{\bf 103} dependencies to be resolved. The result is that the dynamic linker
could attempt nearly 3,600 filesystem operations to resolve the needed
dependencies in the worst case, \emph{every time the process is started}. This
exorbitant cost can be made worse if the store itself resides on a shared
filesystem such as NFS.  This problem is not unique to Nix, and is present in
other store-like systems.  \citeauthor{guix-stat} has written about this problem
for the Guix system~\cite{guix-stat}.  Table~\ref{table:emacs-shrinkwrap} shows the reduction to the \emph{stat} and \emph{openat} syscalls during process startup, and was captured using \emph{strace}. The reduction in syscalls equates to a \textbf{36x} speedup.

\begin{table}[h]
  \begin{center}
    \begin{tabular}{|c| c c||}
      \hline
                    & Calls (stat/openat) & Time (seconds) \\ [0.5ex]
      \hline\hline
      emacs         & 1823                & 0.034121       \\
      \hline
      emacs-wrapped & 104                 & 0.000950       \\
      \hline
    \end{tabular}
  \end{center}
  \caption{Evaluation of emacs stat/openat syscalls}
  \label{table:emacs-shrinkwrap}
\end{table}

\begin{figure}[h]
    \includegraphics[width=\linewidth]{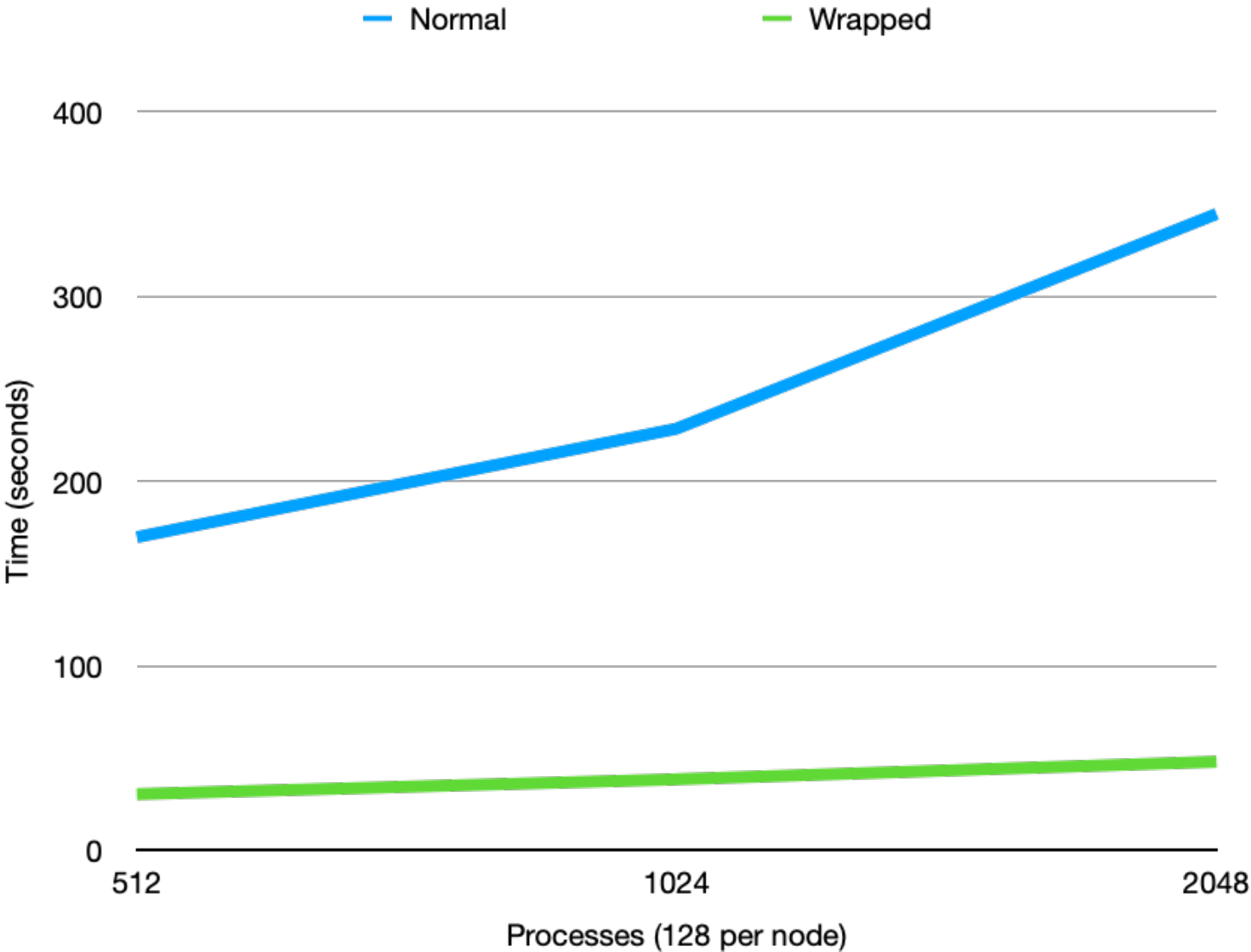}
    \caption{Time-to-launch instances of Pynamic as built (Normal) and shrinkwrapped.}
    \label{fig:shrinkwrap-performance}
\end{figure}

While the total time is not long when considered as part of a single invocation, the effects magnify when applied in the context of even a modestly sized MPI application.  To evaluate larger scale applications we use the Pynamic~\cite{pynamic} dynamic application benchmark to measure the cost to launch and load a large MPI application at a modest scale.  For our experiments, the benchmark is configured to match the general characteristics of a real LLNL application with approximately 900 shared libraries, using the "bigexe" configuration. All modules produced are listed as needed entries on the executable, modified slightly to place each of them in its own rpath directory.  Figure~\ref{fig:shrinkwrap-performance} shows the results of running our Pynamic configuration on a system with two Xeon E5-2695 processors, loading the application and its libraries from NFS.  Each test was run with a cold cache, and negative caching (caching the \emph{non}-existence of a file) is disabled as it is by default on LLNL systems.  At the smallest size, 512 processes on four nodes, the normal executable took 169 seconds to launch, while the wrapped executable took 30.5 for a speedup of $5.5\times$. At 2048 processes, the gap widens to $7.2\times$ for a total \mbox{time-to-launch} of 344.6 seconds for the normal executable.  While this result is on the high end for what can be expected, the costs scale as the scale of the job increases, and the startup time benefits only grow.  Shrinkwrap applies because even though the libraries and Python modules are loaded dynamically by the application, they are known at build time and included in the needed list.  If there were more that were not known, it could be worthwhile to explore combining Shrinkwrap with an approach like Spindle to improve the load performance of those as well.

\subsection{Use Cases}

In the process of preparing for a new AMD-based supercomputer, several software
integration issues have arisen that resist traditional workarounds. The first of these
is caused by a combination of three factors: \Verb`RPATH` entries in the main executable that point to
all of the appropriate libraries, \Verb`LD_LIBRARY_PATH` set in modules to help with internal
library search issues in ROCM packages, and those same ROCM packages using \Verb`RUNPATH` in
place of \Verb`RPATH`. Any one of these would not be a problem by itself, even any two of them,
but all three combined produces unfortunate effects. Specifically, an application built
with ROCM version 4.5 will segfault if run when the module for a different ROCM version
is loaded. This happens because after the first ROCM library is loaded, having been
found by \Verb`RPATH`, the presence of a \Verb`RUNPATH` inside the library causes the loader to ignore
the \Verb`RPATH` entries. The loader then prioritizes the now incorrect \Verb`LD_LIBRARY_PATH`,
causing incorrect versions of the internal libraries used in ROCM to be loaded. Applying
Shrinkwrap and linking all dependencies directly to the binary fixes this issue given a
built binary inside a consistent environment.

The second issue comes from a workaround used inside a vendor library. When using the
system compiler on an El Capitan Early Access system, compiling with OpenMP links in
\Verb|libomp.so|, \emph{without} OpenMP links \Verb|libompstubs.so| instead. This is
perfectly reasonable, it means OpenMP runtime calls are always available. The downsides
are twofold: the application is now dependent on load order to work correctly, and the
linking approach to the Needy Executables workaround \emph{does not work}. The load
order is important because if \Verb|libompstubs.so| loads first the application will run
with no threading or offload support. The workaround breaks because the stub library and
the main OpenMP library are drop-in replacements, and define the same symbols. When both
are loaded at runtime this is fine; whichever loads first wins. When both are specified
on a link line, the link fails due to the duplicates. Since Shrinkwrap does not depend
on manipulating the link line it can encode the required libraries without duplicate
symbol conflicts. Once re-written to absolute paths, the initial load for all needed
libraries is no longer environment dependent and can be inspected in the build
environment and \emph{relied upon} in the user's run environment. Though Shrinkwrap does
not explicitly check symbol shadowing or load orders, it preserves the order the user
set. This prevents a common class of errors in production codes on HPC systems,
especially where multiple compiler stacks must be used and the results linked together.


%% file: text/related.tex
\section{Related Work}

The labor, time, and \emph{art} that goes into producing package distributions for Linux and the software packages that comprise them is not often a subject written about, at least publicly. This research seeks to begin to provide a common set of language and a survey of the current landscape to begin to foster continued discussion.

Recently, there has been a renewed interest in emerging platforms as the landscape of specialized hardware has grown. The quest for portability of software to multiple platforms is in the same effort to better understand the taxonomy of packages, their dependencies, and the ability to reproduce the software artifact. \citeauthor{10.1145/3213846.3213855}'s research\cite{10.1145/3213846.3213855} into how portable software can be given changes to the toolchain (C/C++ compiler) or standard C library demonstrates the challenge of producing reproducible platform-independent software. In their effort to survey the portability of a large corpus of packages, the distribution's \emph{dependency factor} was a constant hurdle. Many popular Linux distributions' packages (e.g., Debian, Fedora) have a large transitive closure, demonstrating how connected the package graph is. Research into the \emph{Redirected Execution Daemon} was necessary to silently bypass failures, allowing builds to progress to completion and avoid the domino effect of failing packages to their upstream dependents.

The focus of our paper with respect to package taxonomies was focused at the lowest runnable unit within a Linux environment, an ELF file (executable or shared object), which often assumes that the language used was C/C++.
This focus helps narrow the differences amongst package management solutions as they differ across distributions.
The cornucopia of languages, however, is much larger than this narrower view, and many come with their own package management solution. Many of these languages reside atop a virtual machine and typically have their own concept of modules and resolution strategies. An interesting field of research is in the ability of this diaspora of language package management tooling to communicate amongst each other, potentially with that of the operating system, and avoid duplicating efforts\cite{10.1145/3365137.3365402}.


%% file: text/conclusion.tex
\section{Conclusions}

Package dependencies in HPC and across the ecosystem have increased drastically in recent years. As the dependencies between software have become increasingly interconnected,
package management solutions have emerged to address new deployment models. In
this paper, we surveyed the most well-known deployment models in use today in
Unix-like environments and identified challenges each face as they make certain
trade-offs for desired guarantees. 

Distributions and package managers have begun to make leaps into greater atomicity, security, and
reproducibility, but these goals are coming into conflict with the
semantics of the ELF loader and incurring non-trivial costs as a result.  We
have presented some solutions that work around many of the common problems in
this space within those options, and presented our tool for mapping out the chaos of dependencies: Shrinkwrap. Shrinkwrap is our solution to some of the common issues found in composing packages across models and from different sources into a reliable executable.  In doing so, we have shown we can improve load times for some highly dynamic applications by a factor of seven, while making the application \emph{easier} to launch for users.

%% file: text/acknowledgements.tex
\section{Acknowledgements}

This work was in part supported by the National Science Foundation under Cooperative Agreement OAC-1836650, by the US Department of Energy ASCR DE-NA0003525 (FWP 20-023266, subcontract with Sandia National Labs), and by the Center for Research in Open Source Software (\url{https://cross.ucsc.edu}).  This work was performed under the auspices of the U.S. Department of Energy by Lawrence Livermore National Laboratory under contract DE-AC52-07NA27344. Lawrence Livermore National Security, LLC.